\documentclass[11pt,a4paper]{article}
\usepackage[utf8]{inputenc}
\usepackage{amsmath}
\usepackage{amsfonts}
\usepackage{amssymb}
\usepackage{geometry}
\usepackage{slashed}
\begin{document}
\begin{center}
{\Large \textbf{DUALITY AND DILATON}}

\vspace{0.6cm}

\textbf{A. A. Tseytlin}\footnote{On leave of absence from the Department of Theoretical Physics,
P. N. Lebedev Physics Institute, Moscow 117924, USSR.
Supported in part by the National Science Foundation, grant No. PHY90-96198.
}

\renewcommand{\thefootnote}{{footnote}}

\vspace{0.2cm}

Department of Physics and Astronomy, The Johns Hopkins University, Baltimore,\\
MD 21218, USA

\vspace{0.5cm}

%Received 6 May 1991

{\it Published in Mod. Phys. Lett. A6 (1991) 1721-1732.}

\end{center}

\vspace{0.3cm}

{\small  We review and elaborate on the issue of the dilaton transformation under the usual
$r \rightarrow \alpha'/r$ target space duality and its ``non-static'' generalization
(or $\sigma$-model duality). It is found that the transformation law
$r \rightarrow \alpha'/r$,\  $\phi \rightarrow \phi - \ln(r/\sqrt{\alpha'})$
which guarantees duality at the one-loop $\sigma$-model level should be modified
at two (and higher) loop order. The ``non-static'' duality is illustrated on the
example of ``cosmological'' solutions in $D \ge 2$ with time-dependent radii of
space torus.}

\vspace{1.2 cm}

\setcounter{footnote}{0}

\def \foot   {\footnote}  \def \no {\nonumber}

\noindent
\textbf{1.} Target space duality was originally discovered $[1]$ as a symmetry of
the spectrum of states and (hence) of the one-loop partition function of the closed
string theory compactified on a torus,

\[
M_{nm}^2(r) = \frac{n^2}{r^2} + \frac{m^2 r^2}{\alpha'^2} + \ldots
= M_{mn}^2(\tilde r), \qquad
Z_1(r) = Z_1(\tilde r), \qquad
\tilde r = \frac{\alpha'}{r}.
\tag{1}
\]
Since the duality between the momentum and winding modes implies the symmetry
of the corresponding set of vertex operators, the duality should appear to be the
symmetry of the full string theory (scattering amplitudes, etc.) expanded near the
torus vacuum  [2]. As was first pointed out in Ref.~3, for this to be actually true
the dilaton $\phi$ (or the string coupling constant $g = \exp \phi_{\text{vac}}$)
should also transform under the duality
\[
\tilde a = a^{-1}, \qquad
\tilde \phi = \phi - \ln a, \qquad
a \equiv \frac{r}{\sqrt{\alpha'}} .
\tag{2}
\]
Here we  consider a  circle  of dimensionless radius $a$.
In particular, it is (2) that is the symmetry of the higher loop
(genus $n \ge 1$) vacuum partition function,
\[
Z_n(g,a) = Z_n(\tilde g,\tilde a), \qquad
\tilde g = \frac{g}{a}, \qquad
Z_n = g^{2(n-1)} \tilde Z_n(a).
\tag{3}
\]
The necessity of the constant dilaton shift for the duality invariance of the string
partition function was also noted in Ref.~4. The symmetry (3) was demonstrated
explicitly e.g.\ in Refs.~5 and 6. 

The reason why the dilaton should be shifted
under the duality is easy to understand from a field theoretic point of view  [3]. 
Consider a string field theory action,
\[
S=\int d^D x\, e^{-2\phi}(\ldots)
   =\frac{1}{g^2}\int dy \int d^{D-1}x
   \Big(\varphi \Delta \varphi + \varphi^3 + \ldots \Big),
\]
where $y$ is the periodic coordinate. Integrating over $y$ we get the factor
of $a$ in front of the action for the Fourier modes. The theory will depend on
$a$ only through the effective ``$(D-1)$-dimensional'' coupling
$\bar g = g/\sqrt{a}$ and the masses of fields. Since the latter are symmetric
under $a \to a^{-1}$ (cf.\ (1)) $\bar g$ should not change under the duality
in order for the theory to be duality invariant. The invariance of $\bar g$
implies that $g$ should transform according to (3).

Recently, an attempt was made to understand if there exists a generalization of
the duality to the case of ``non-static'' string vacua for which the radius of
the torus may depend on other coordinates $x^\mu$ (e.g.\ on time)  [7].
 As was
emphasized in Ref.~8, one should expect to find duality as a symmetry between
string vacua with ``radii'' $a$ and $a^{-1}$ since for any $x$-dependent
perturbation of a ``static'' theory (compactified on a torus) there is a
corresponding one in the dual theory (compactified on a ``dual'' torus).\foot{In the ``static'' case one can view duality either as a symmetry of a given
conformal theory or as a symmetry which relates two conformal theories. The
former interpretation, though less obvious in the ``non-static'' case, should be
possible once the analogs of the usual winding modes are identified. } 
It was found  [8]  that the exact ``plane wave'' classical solutions  [9]  of the
bosonic string theory discussed in Ref.~7 have dual analogs provided one
transforms the dilaton according to (2) with $a$ now being $x$-dependent.

The transformation between the theories with ``radii'' $a(x)$ and $a^{-1}(x)$
can be interpreted as a $\sigma$-model duality transformation for the
corresponding string $\sigma$-models. The need to shift the dilaton field in
order for the duality related $\sigma$-models to have equivalent one-loop Weyl
anomaly coefficients (``$\bar \beta$-functions'') was, in fact, already
pointed out in Ref.~10,
\[
\tilde{\phi}=\phi-\frac{1}{2}\ln G_{11},
\qquad
\tilde{G}_{11}=G_{11}^{-1},
\qquad
G_{11}=a^2(x).
\tag{4}
\]
Below we shall first rederive the result of Ref.~10 in a simple way, explaining
why the one-loop $\bar {\beta}$-functions (Weyl anomaly coefficients) 
 and the leading term in the effective
action are invariant under (4). We shall then determine how the duality
transformation law (4) should be modified at the two-loop level so that the
couplings of the ``dual'' $\sigma$-models be solutions of the two-loop conformal
invariance conditions. Though a possibility of a modification of (4) by
$O(\alpha')$ terms seems to be anticipated before  [8,10]  we shall see that
there is no \emph{a priori} reason to expect these additional terms to be local.
Still, the modification necessary at the two-loop level turns out to be local.
We shall find that the effective action is invariant under a generalization of
(4) but, in contrast to the one-loop case, one of the two-loop
$\bar{\beta}$-functions is no longer duality invariant ``off-shell'', i.e.,
it is only the solutions of the conformal invariance conditions that are
related by the duality transformation. Finally, we shall discuss some
cosmological solutions.

\bigskip

\bigskip

\noindent
\textbf{2.} We shall start with the following $\sigma$-model $(\mu,\nu = 1,\ldots,D)$
\[
I = \frac{1}{4\pi\alpha'} \int d^2 z \sqrt{g}
\Big[\hat{G}_{\mu\nu}(X)\partial_a X^\mu \partial^a X^\nu
+ \alpha' R^{(2)}\phi(X)\Big].
\tag{5}
\]
We shall ignore the antisymmetric tensor coupling which is not important for the
issue of the dilaton transformation under duality.
 One is able to define the
``dual'' $\sigma$-model if $\hat{G}$ and $\phi$ are independent of a number of
coordinates (i.e.\ if the metric has a number of Killing vectors, etc.). We shall
consider the simplest case of one isometry and assume that the metric has a
``block-diagonal'' form
\[
\hat{G}_{11}=a^2(x)=e^{2\lambda(x)}, \qquad
\hat{G}_{1i}=0, \qquad
\hat{G}_{ij}=G_{ij}(x), \qquad
\phi=\phi(x), \nonumber 
\]
\[
X^1=y, \qquad X^i=x^i, \qquad i=2,\ldots,D .
\tag{6}
\]
The coordinate $y$ corresponds to a periodic coordinate in the string duality
context (the topology of $y$ is not important in the $\sigma$-model duality
context, i.e., for a discussion of local objects like $\bar \beta$-functions).
The metrics considered in Refs.~7 and 8 can be represented as particular cases
of the metric (6). The $\sigma$-model (5),(6), i.e. 
\[
I = \frac{1}{4\pi\alpha'} \int d^2 z \sqrt{g}
\Big[G_{11}(x)\partial_a y \partial^a y
+ G_{ij}(x)\partial_a x^i \partial^a x^j
+ \alpha' R^{(2)}\phi(x)\Big]
\tag{7}
\]
is related to the ``dual'' one by the standard chain of formal transformations
(we use Euclidean signature)
\[
\frac{1}{2}G_{11}\partial_a y \partial^a y\ \ 
\rightarrow\ \ 
\frac{1}{2}G_{11}^{-1}p_a p^a + i p^a \partial_a y\qquad 
\rightarrow\qquad 
p^a = \epsilon^{ab}\partial_b \tilde{y}\ \ 
\rightarrow\ \ 
\frac{1}{2}G_{11}^{-1}\partial_a \tilde{y}\partial^a \tilde{y},
\tag{8}
\]
i.e., the model which is classically equivalent to (7) is
\[\no
\tilde{I} =
\frac{1}{4\pi\alpha'} \int d^2 z \sqrt{g}
\Big[\tilde{G}_{11}(x)\partial_a \tilde{y}\partial^a \tilde{y}
+ \tilde{G}_{ij}(x)\partial_a x^i \partial^a x^j
+ \alpha' R^{(2)}\tilde{\phi}(x)\Big],\no 
\]
\[
\tilde{G}_{11}=G_{11}^{-1}, \qquad
\tilde{G}_{ij}=G_{ij}, \qquad
\tilde{\phi}=\phi .
\tag{9}
\]
The question we would like to address is about an equivalence of the models
(7) and (9) at the quantum level. Since the quantum analog of the transformation
(8) produces non-trivial ($x$-dependent, non-local) ``Jacobian'' (Eq.~(13) below)
which appears under the functional integral over $x$ it is not clear why the two
theories (7) and (9) should be equivalent at the quantum level even if one admits
a possibility that the couplings in (9) should be replaced by
\[\tag{10}
\tilde{G}_{11}=G_{11}^{-1}+A, \qquad
\tilde{G}_{ij}=G_{ij}+H_{ij}, \qquad
\tilde{\phi}=\phi-\frac{1}{2}\ln G_{11}+B,
\]
\[
A,B,H = O(\alpha').
\]
$(A,B,H)$ depend on the derivatives of $G_{11}$ so that (10) reduces to the
standard form of the duality transformation (4) in the ``static'' limit.

In fact, the equivalence
between (7) and (9), (10) appears to be true only in the following weak sense:
given a solution $(G_{11},G_{ij},\phi)$ of the Weyl invariance conditions
$(\bar \beta=0)$ there exists another solution represented by the duality
transformations (10) with $A,B,H$ being local functions of $G_{11},G_{ij},\phi$.
Though we do not know how to prove this statement in general, we shall
demonstrate its validity in the two-loop approximation. Moreover, we shall
find that $A$ can be chosen to be zero, so that the original form of the
duality transformation for the ``radius'' $a\rightarrow a^{-1}$ is preserved
at the higher loop level.\foot{As we shall see, Eq.~(10) with $A,B,H=0$ gives a relation between the
two solutions of the one-loop Weyl invariance conditions [10]. 
 Since the
exact solution of $\bar \beta=0$ can always be represented as a one-loop
solution plus a series in $\alpha'$ one may argue that the transformation
(10) in general maps a solution into a solution. The non-trivial questions
are why $A,B,H$ should be local and why $A$ can be chosen to be zero. }

To compare the quantum theories corresponding to the actions (7) and (9)
one should, in general, introduce sources (or backgrounds) for $y,\tilde y,$
and $x$. Let us ignore the backgrounds for $y$ and $\tilde y$, integrate these
fields out and compare the resulting functional integrals over $x$.
Integration over $y$ produces the following ``effective action'' for $x$
\[
\hat I =
\frac{1}{4\pi\alpha'}\int d^2z \sqrt{g}
\Big[G_{ij}(x)\partial_a x^i \partial^a x^j
+ \alpha' R^{(2)}\phi(x)\Big] + W[x,\lambda],
\tag{11}
\]
\[
W = \tfrac{1}{2}\ln \det \Delta , \qquad
\Delta(\lambda) = -\nabla^a\big(e^{2\lambda(x)}\nabla_a\big).
\tag{12}
\]
Integration over $\tilde y$ in (9) gives (11) with $W[x,\lambda]$ replaced
by $W[x,-\lambda]$ (we do not indicate explicitly the dependence of $W$ on
$g_{ab}$). The difference between the ``original'' and ``dual'' effective
theories for $x$ is thus given by
\[
\omega[x,\lambda] = W[x,\lambda] - W[x,-\lambda].
\tag{13}
\]
The reason why the duality transformation (8) produces the non-trivial
``Jacobian'' $\omega$ (13) is that since the auxiliary field $p_a$ couples
to the derivative of $y$ (and its ``transverse'' part does not decouple for
non-constant $\lambda$) the Gaussian functional integral over $p_a$ should
actually be understood as the integral over the two scalar fields,
$p_a=\partial_a f+\epsilon_a{}^b\partial_b h,\;
\epsilon_{ab}=\varepsilon_{ab}\sqrt{g}$. Therefore, $\omega$ has the
following functional representation (an equivalent representation was used
in Ref.~10)
\[
\exp(-\omega[x,\lambda]) =
N\int [df\,dh]\exp(-I[f,h]),
\tag{13'}
\]
\[
I[f,h] =
\tfrac{1}{2}\int d^2z \sqrt{g}\,
e^{2\lambda}\big(\partial_a f + \epsilon_a{}^b \partial_b h\big)^2,
\]
where $N$ is a $\lambda$-independent factor. In conformal coordinates
\[
I[f,h] =
2\int d^2z\, e^{2\lambda}\,\partial F\,\bar{\partial}\bar{F},
\qquad
F=f+ih, \qquad
\partial = \tfrac12(\partial_1-i\partial_2),
\tag{13''}
\]
\[
\omega = \tfrac12 \ln \det Q, \qquad\qquad 
Q = -\partial e^{2\lambda}\bar{\partial}.
\]
$\omega$ is a complicated functional of $\lambda$ but its Weyl anomaly
part can be computed exactly (see Eq.~(18) below). Let us note that for
the actual computation of $\omega$ the
representation (13$'$) is less useful than (13).\foot{It may seem that the $\lambda$-dependence 
of the determinant of $Q$ can be
found explicitly by using the same trick as in the computation of curved space
Laplacians in the conformal gauge. Namely,
\[
\delta\omega
=
\frac12\int_{\epsilon}^{\infty} {ds\over s}\,
\mathrm{tr}(2s\,\partial \delta\lambda\,e^{2\lambda}\bar{\partial}\exp(-sQ))
=
-\int_{\epsilon}^{\infty} ds\,
\mathrm{tr}(\delta\lambda\,\tilde {Q}\exp(-s\tilde Q))
=
\mathrm{tr}(\delta\lambda\,\exp(-\epsilon\tilde Q)),
\]
where $\tilde Q=-e^{2\lambda}\partial\bar{\partial}$ and we have used
$Be^{AB}=e^{BA}B$. Given the standard expression for the heat kernel of
$\tilde Q$,
\[
\mathrm{tr}\,e^{-\epsilon\tilde Q}
\simeq
\frac{1}{4\pi\epsilon} + \frac{1}{12\pi}\partial^2\lambda + \ldots,
\]
one would have to conclude that $\omega \sim \int d^2z\,\partial\lambda
\bar{\partial}\lambda$. This is definitely a wrong result since according to
(13) $\omega$ should change sign under $\lambda\rightarrow-\lambda$. It is
possible to check directly that there is a non-vanishing $O(\lambda^3)$ term
in $W$ and $\omega$. The reason for the failure of the above formal manipulations
is that $Q$ cannot be represented as a product of two operators which are
adjoint to each other, i.e., that $\tilde Q$ is not a self-adjoint operator (the
norms in the functional spaces, which in the case of the curved space
Laplacians depend on the conformal factor for covariance, here are
$\lambda$-independent).
  } 

In defining the quantum theories for the $\sigma$-models (7) and (9) one should
make a choice of a regularization and a functional measure. The basic principle
is that one should guarantee the covariance of the functional integral in the
full $D$-dimensional space (after all, (6) is just a particular background in the
original $D$-dimensional $\sigma$-model (5)). One should use the same
regularization for the action and the measure (cf.\ Ref.~11) and also one and
the same regularization for all the fields in the theory ($y,x,$ and $\tilde y$).
We shall employ the dimensional regularization which is manifestly covariant and
in which one can disregard Jacobians of local field redefinitions (since
$\delta^{(2)}(z,z)=0$). In particular, one can ignore the contribution of the
local measure (except for its zero mode part).\foot{Results in different regularization schemes are related by
redefinitions of couplings (see, e.g., Ref.~12 for a discussion of different
regularizations). Finite (non-covariant) coupling definitions may also be
necessary if a choice of a regularization or measure does not respect
covariance  [11]. Under the choice of the measure and regularization which
guarantees the covariance of the $\sigma$-model perturbation theory all power
divergences cancel out; this implies, in particular, that the tachyon coupling
does not participate in the $\sigma$-model duality transformation.}

To compute $W$ (12) in dimensional regularization it is more convenient to use
its equivalent representation which is found if one first makes the rescaling
$y = e^{-\lambda}y'$ and then integrates over $y'$ in (7) (note that
$
\int d^2z\sqrt{g}(\partial_a y' + A_a y')^2
=
\int d^2z\sqrt{g}\Big[(\partial_a y')^2 + V y'^2\Big],
\ \ 
V = -\nabla^a A_a + A_a A^a $)
\[
W = \tfrac{1}{2}\ln\det\hat{\Delta},
\qquad\qquad
\hat{\Delta} = -\nabla^2 + V ,
\tag{14}
\]
\[
V = \nabla^2\lambda + \partial^a\lambda\,\partial_a\lambda,
\qquad\qquad 
\partial_a\lambda = \partial_a x^i\,\partial_i\lambda .
\tag{15}
\]
Observing that $\int d^2z\sqrt{g}\, V y'^2$ is classically Weyl invariant,
the anomaly is easy to compute in dimensional regularization  [13]
\[
\frac{2g_{ab}}{\sqrt{g}}\frac{\delta W}{\delta g_{ab}}
=
\frac{1}{4\pi}b_2,
\qquad\qquad 
b_2 = \frac{1}{6}R^{(2)} - V .
\tag{16}
\]
%%%%%%%%%%%%%%%%%%%%%%%%%%%%%%%%%%%%
Integrating the anomaly we find\foot{A simple way to check the coefficient of the $R^{(2)}\lambda$ term in
$W$ is to start with $\frac{1}{4\pi}\int d^2z\sqrt{g}e^{2\lambda}
\partial_a y\partial^a y$, expand in powers of $\lambda$ and to use that
$\langle \partial_a y\partial^a y\rangle = -\frac{1}{2}R^{(2)}$ in
dimensional regularization [12]. 
 The computation of the conformal factor
dependence of the determinant (14) is also straightforward using the
proper-time representation: since the operator in (14) depends on the
conformal factor in the same way as the ``free'' Laplacian, the conformal
variation of its determinant can be expressed in terms of the heat kernel
coefficient (16).} 
\[ W[x,\lambda] =
\frac{1}{4\pi}\ln\epsilon
\int d^2z\sqrt{g}\big(\frac{1}{6}R^{(2)}-\nabla^2\lambda
-\partial^a\lambda\,\partial_a\lambda\big)
-\frac{1}{96\pi}\int R^{(2)}\nabla^{-2}R^{(2)}
\]
\begin{equation}
\qquad \qquad \qquad \qquad -\frac{1}{8\pi}\int d^2z\sqrt{g}
(\nabla^2\lambda+\partial_a\lambda\partial^a\lambda)\nabla^{-2}R^{(2)}
+\tilde W[x,\lambda],
\qquad\ \ \  
\ln\epsilon=\frac{1}{d-2} ,
\tag{17}
\end{equation}
where $\tilde W$ is the non-trivial part of $W$ which does not depend on the
conformal factor of the metric. Since it is only the terms which are of odd
power in $\lambda$ that may be present in $\omega$ (13), we find (after
integrating by parts)
\[
\omega =
-\frac{1}{4\pi}\int d^2z\sqrt{g}R^{(2)}\lambda(x)
+\tilde\omega[x,\lambda],
\qquad\qquad \ \ \ 
\tilde\omega =
\tilde W[x,\lambda]-\tilde W[x,-\lambda].
\tag{18}
\]
We conclude that in order to preserve the equivalence of (7) and (9) at least
in the one-loop approximation one is to shift the dilaton in (9)  [10]
\begin{equation}
\tilde{\phi}=\phi-\frac{1}{2}\ln G_{11}=\phi-\lambda\ ,
\qquad\qquad \ \ \ 
\tilde{\lambda}=-\lambda .
\tag{19}
\end{equation}
In the above derivation of (18) it was assumed that $\lambda(x)\neq const$.
One may question if there is a connection between the transformation law (19)
and the constant shift (2). From the $\sigma$-model point of view, the latter
is suggested by the presence of the measure factor
$\sqrt{\hat G}=e^\lambda\sqrt{G}$ in the partition function (and effective
action). Suppose the model (7) is defined on a compact 2-space with the
topology of a sphere. Then the partition function contains the zero mode
factor  [14]
\[
\int dy_0\int d^{D-1}x_0\sqrt{G}e^\lambda e^{-2\phi}
\]
(other parts of the formal measure $\Pi_z\sqrt{\hat G(x(z))}$ do not
contribute in dimensional regularization). The corresponding factor for the
``dual'' model (9) is
\[
\int d\tilde y_0\int d^{D-1}x_0\sqrt{G}e^{-\lambda}e^{-2\tilde\phi}.
\]
The two factors are equivalent if we use (19) (and ignore the integrals over
$y_0$ and $\tilde y_0$ which decouple).

\bigskip

\noindent
\textbf{3.}
As we have already mentioned, the complicated structure of $\omega$ in (18)
makes it unlikely that the quantum properties of the models (7) and (9),
(19) will, in general, be the same at higher than one-loop order. In
particular, computing the integral over $x$ one finds that $\tilde\omega$
provides additional contributions to the $\beta$-function of $G_{ij}$.

To compare the Weyl anomaly coefficients corresponding to the models (7)
and (9) it is not actually necessary to go through direct computations:
one may start with the general expressions known for the $\sigma$-model
(5) and specify to the particular background (6), i.e., to the models
(7),(9). The two-loop Weyl anomaly coefficients
%%%%%%%%%%%%%%%%%%%%%%%%%%%%%%%%%%%%%%%%
for the $\sigma$-model (5) (computed using dimensional regularization and
normal coordinate expansion corresponding to a nonlinear redefinition of
the quantum field) are given by  [15,16] 
\[
\bar{\beta}_{\mu\nu}
=
\hat{R}_{\mu\nu}
+2\hat{D}_\mu \hat{D}_\nu \phi
+\frac{1}{2}\alpha'\hat{R}_{\mu\alpha\beta\delta}
\hat{R}_\nu{}^{\alpha\beta\delta}
\equiv
\bar{\beta}^{(1)}_{\mu\nu}
+\alpha'\bar{\beta}^{(2)}_{\mu\nu},
\tag{20}
\]
\[
\bar{\beta}^{\phi}
=
\frac{1}{6}(D-26)
-\frac{1}{2}\alpha'\hat{D}^2\phi
+\alpha'(\partial\phi)^2
+\frac{1}{16}\alpha'^2
\hat{R}^2_{\lambda\mu\nu\rho}\ .
\]
It is sufficient to consider only $\bar{\beta}_{\mu\nu}$ since the structure
of $\bar{\beta}^{\phi}$ is determined by that of $\bar{\beta}_{\mu\nu}$.
The following components of the connection and curvature are non-vanishing
for the background (6)
\begin{align} 
&\no ds^2 = e^{2\lambda(x)}dy^2 + G_{ij}(x)dx^i dx^j ,
\\
& \no \hat{\Gamma}^{\,i}_{jk}=\Gamma^{\,i}_{jk}, \qquad
\qquad
\hat{\Gamma}^{\,1}_{i1}=b_i, \qquad 
\hat{\Gamma}^{\,i}_{11}=-e^{2\lambda}b^{\,i}, \qquad
b_i \equiv \partial_i\lambda,
\no \\   &
\hat{R}_{ijkl}=R_{ijkl}, \qquad
\hat{R}_{1i1j}=-e^{2\lambda}b_{ij}, \qquad
b_{ij}\equiv D_i b_j + b_i b_j,
\no \\  &
\hat{R}_{11}=-e^{2\lambda}b^{\,i}{}_i, \qquad
\hat{R}_{ij}=R_{ij}-b_{ij}, \qquad
\hat{R}=R-2b^{\,i}{}_i,
\no \\  &
\hat{D}_i\hat{D}_j\phi=D_iD_j\phi, \qquad
\hat{D}_1\hat{D}_1\phi=e^{2\lambda}b^{\,i}\partial_i\phi .
\tag{21}
\end{align}
Since all the ``mixed'' $(1i)$-components of the second rank tensors
vanish, $\bar{\beta}_{1i}=0$, the  one-loop $\bar{\beta}$-functions for the
model (7) are thus given b
\[
\bar{\beta}^{(1)}_{11}
=
- e^{2\lambda}
\Big(
D_iD^i\lambda
+ \partial_i\lambda\,\partial^i\lambda
-2\partial^i\lambda\,\partial_i\phi
\Big),
\tag{22}
\]
\[
\bar{\beta}^{(1)}_{ij}
=
R_{ij}
- D_iD_j\lambda
- \partial_i\lambda\,\partial_j\lambda
+2D_iD_j\phi .
\tag{23}
\]
The corresponding functions in the dual theory (9) are given by (22), (23)
with $\lambda \rightarrow \tilde{\lambda}=-\lambda$, $\phi\rightarrow
\tilde{\phi}=\phi-\lambda$. It is easy to see that this ``duality
transformation'' is the invariance of the one-loop $\bar{\beta}$-functions
(22), (23) (note that $\beta_{11}=\frac{dG_{11}}{d\ln\epsilon}$,
$G_{11}=e^{2\lambda}$)
\[
\bar{\beta}^{(1)}_{11}e^{-2\lambda}
=
-\tilde{\bar{\beta}}^{(1)}_{11}e^{-2\tilde{\lambda}},
\qquad
\bar{\beta}^{(1)}_{ij}
=
\tilde{\bar{\beta}}^{(1)}_{ij}.
\tag{24}
\]
This, of course, should have been expected in view of the above discussion
(cf.\ (18), (19)). The transformation (19) is also the invariance of the
leading term in the effective action (which corresponds to
$\bar{\beta}^{(1)}_{\mu\nu}$)
%%%%%%%%%%%%%%%%%%%%%%%%%YYYYYYY
\[
\hat{S}
=
\int d^D X \sqrt{\hat{G}}\,e^{-2\phi}
\Big[
c
+\alpha'(\hat{R}+4\partial_\mu\phi\partial^\mu\phi)
+\frac{1}{4}\alpha'^2
\hat{R}^2_{\lambda\mu\nu\rho}
+\ldots
\Big],
\qquad
c=\tfrac{2}{3}(26-D),
\tag{25}
\]
\[
\hat{S}
\rightarrow
S^{(1)}[\lambda,\phi,G]
=
\int d^{D-1}x\sqrt{G}e^{\lambda}e^{-2\phi}
\Big[
c
+\alpha'
\Big(
R
-2D^iD_i\lambda
-2\partial_i\lambda\partial^i\lambda
+4\partial_i\phi\partial^i\phi
\Big)
\Big]
\]
\[
\ \ \ \ =
\int d^{D-1}x\sqrt{G}
e^{\lambda-\phi}e^{-\phi}
\Big[
c
+\alpha'(R-4\partial_i(\lambda-\phi)\partial^i\phi)
\Big],
\tag{26}
\]
\[
S^{(1)}[\lambda,\phi,G]
=
S^{(1)}[-\lambda,\phi-\lambda,G].
\tag{27}
\]
Note that since the action (26) contains derivatives of $\lambda$ and $\phi$
the remark that the measure factor $e^\lambda e^{-2\phi}$ is invariant under
(19) is not by itself sufficient to prove the (``one-loop'') duality
invariance (27) (one should either go through the general argument based on
computation of $\omega$ (13), (18) or do the explicit check; see also below).

Let us now consider the two-loop approximation. Using (21) we get from
(20), (25) and (26)
\[
\bar{\beta}^{(2)}_{11}
=
e^{2\lambda}\Big(D_i D_j \lambda + \partial_i \lambda
\partial_j \lambda \Big)^2 ,
\tag{28}
\]
\[
\bar{\beta}^{(2)}_{ij}
=
\frac{1}{2} R_{ikln} R_{j}{}^{kln}
+
\big(D_i D_k \lambda + \partial_i \lambda \partial_k \lambda\big)
\big(D_j D^k \lambda + \partial_j \lambda \partial^k \lambda\big),
\]
\[
S^{(2)}
=
\int d^{D-1}x \sqrt{G}\, e^\lambda e^{-2\phi}
\Big[
\frac{1}{4}R_{ijkl}^2
+
\big(D_i D_j \lambda + \partial_i \lambda \partial_j \lambda\Big)^2
\big].
\tag{29}
\]
It is straightforward to check that the following transformation leaves
the $\bar{\beta}$-function for $G_{ij}$,
$\bar{\beta}_{ij}=\bar{\beta}^{(1)}_{ij}+\alpha' \bar{\beta}^{(2)}_{ij}$,
and the effective action $S=S^{(1)}+\alpha' S^{(2)}$ invariant for
arbitrary values of the couplings (cf.\ (10))
\[
\tilde{\lambda}
=
-\lambda+\eta ,
\qquad
\tilde{\phi}
=
\phi-\lambda+\frac{1}{2}\eta ,
\qquad
\tilde{G}_{ij}=G_{ij},
\qquad
\eta=\alpha'(\partial_i \lambda)^2 .
\tag{30}
\]
Note that (30) is the symmetry of the measure factor $e^\lambda e^{-2\phi}$
in $S$ for arbitrary $\eta$. The action is invariant because the variation
of the $\partial_i(\lambda-\phi)\partial^i\phi$ term in (26) changes the
sign of the $O(\lambda^3)$ term in the transformed form of (29). It seems
likely that a similar mechanism will provide the invariance of the action
at higher orders in $\alpha'$. This is suggested, in particular, by the
existence of the following representation for the effective action  [17]
\[
\hat{S}
=
-\frac{1}{2}\Big(\frac{\partial}{\partial t}
\int d^D X \sqrt{\hat{G}} e^{-2\phi}\Big)_{t=0}
=
\int d^D X \sqrt{\hat{G}} e^{-2\phi}
\Big(\bar{\beta}^\phi-\frac{1}{4}\bar{\beta}^{\hat{G}}\Big),
\]
\[
\hat{S}\rightarrow
S
=
-\frac{1}{2}\Big(\frac{\partial}{\partial t}
\int d^{D-1}x \sqrt{G} e^\lambda e^{-2\phi}\Big)_{t=0}.
\]
Here $t$ is the logarithm of the cutoff, the couplings are assumed to be
the bare ones and we have specified the general expression to the case
of the background (21). Given that (30) is the invariance of the measure
factor one may thus expect (30) to be also the invariance of the full
action. This observation is only a hint since it does not seem to rely
upon a particular choice of $\eta$ in (30) (the above representation for
$\hat{S}$ is valid only for a particular definition of the couplings,
i.e., in a particular renormalization scheme).

The invariance of the effective action does not, however, imply the
invariance of the corresponding Weyl anomaly coefficients: the
``off-shell'' relation
\[
\frac{\delta S}{\delta \varphi^A}
=
\kappa_{AB}\, \bar{\beta}^B ,
\qquad
\varphi^A=(\hat{G}_{\mu\nu},\phi),
\]
involves the field dependent matrix $\kappa_{AB}$ which is non-trivial
in the two-loop approximation (it contains $O(\alpha')$ terms)   [16] 
and is not invariant under (30). In fact, the transformation (30) is
not a symmetry of $\bar{\beta}_{11}$. It is the symmetry only if the
couplings satisfy $\bar{\beta}_{\mu\nu}=0$ so that we are permitted
to use the leading order equations $\bar{\beta}^{(1)}_{\mu\nu}=0$
(22), (23) to get rid of the dilaton term
$\partial^i(\partial\lambda)^2 \partial_i \phi$.

Therefore, in contrast to the one-loop case, there is no local transformation
of the couplings of the form (10) that is the symmetry of the two (and higher)
loop $\bar{\beta}$-functions themselves (as functions of arbitrary couplings).\foot{To prove this let us make the general transformation
$\tilde{\lambda}=-\lambda+\eta$, $\tilde{\phi}=\phi-\lambda+B$,
$\tilde{G}_{ij}=G_{ij}+H_{ij}$, where $\eta$, $B$, and $H$ do not depend on
$\phi$ (note that the duality transformation in the functional integral and
in particular $\omega$ in (18) does  not involve $\phi$). We find that both
$\eta$ and $H$ should be non-vanishing in order to cancel the extra
$\phi$-dependent $O(\alpha')$ term in $\bar{\beta}_{11}$. At the same time,
$H$ should vanish in order to avoid similar $\phi$-dependent $O(\alpha')$
term in $\bar{\beta}_{ij}$. The transformation with only $B$ being
non-vanishing does not leave $\bar{\beta}$-functions invariant either.  } 

The presence of the $\alpha'$-correction in $\tilde{\lambda}$ in (30) may
look undesirable since it modifies the usual $(a(x)\rightarrow a^{-1}(x))$
duality transformation law. It is possible to preserve the relation
$\tilde{\lambda}=-\lambda$ and yet get a symmetry of the solutions of
$\bar{\beta}_{\mu\nu}=0$. The set of solutions is invariant under the
transformations of the couplings induced by the general coordinate
transformation of $x$: $\delta\lambda=\xi^i\partial_i\lambda$,
$\delta\phi=\xi^i\partial_i\phi$,
$\delta G_{ij}=D_i\xi_j+D_j\xi_i$. Choosing $\xi_i=\alpha'\partial_i\lambda$
and transforming the couplings on the right-hand side of Eqs.~(30) we find
another symmetry transformation
\[
\tilde{\lambda}=-\lambda,
\qquad
\tilde{\phi}
=\phi-\lambda
+\alpha'\Big[\partial^i\lambda\partial_i\phi
-\frac{1}{2}(\partial_i\lambda)^2\Big]
=\phi-\lambda+\frac{1}{2}\alpha' D^2\lambda ,
\tag{31}
\]
\[
\tilde{G}_{ij}=G_{ij}+2\alpha' D_iD_j\lambda .
\]
We have further simplified the $\alpha'$-term in $\tilde{\phi}$ by using the
leading order equation $\bar{\beta}^{(1)}_{11}=0$ (22).

We expect that similar duality transformation relating solutions of the Weyl
invariance conditions will exist at higher loop orders with
$\tilde{\lambda}=-\lambda$ being unmodified and higher order $\alpha'^n$-corrections
to $\tilde{\phi}$ and $\tilde{G}$ being local. It is clear that the shift
$\phi\rightarrow\phi-\lambda$ plays a more fundamental role than
$\alpha'$-corrections to $\tilde{\phi}$ and $\tilde{G}$. In fact, as we have
seen, the transformation (19) is the ``off shell'' symmetry of the one-loop
$\bar{\beta}$-functions [10] and hence it becomes exact once higher loop
corrections are absent (as it is the case for the models considered in
Refs.~7--9).

\

\bigskip

\noindent
\textbf{4.}
The duality (19), (30) may have interesting implications for string
cosmology, relating solutions corresponding to expanding and contracting
universes with different values of the effective string coupling. This
``non-static'' duality can be illustrated explicitly on the example of the
cosmological solution of the string effective equations (Weyl invariance
conditions) (20) found in the leading order approximation in $\alpha'$
in Ref.~18. The solution is represented by the following background
\[
ds^2=-dt^2+\sum_{n=1}^{D-1}a_n^2(t)\,dx_n^2 ,
\tag{32}
\]
\[
\qquad \ \ a_n(t)=c_n(\tanh bt)^{p_n},
\qquad 
n=1,\ldots,D-1,
\]
\[
\phi(t)=\phi_0-\ln\cosh bt+\gamma\ln\tanh bt .
\tag{33}
\]
where the constants $b$, $p_n$, and $\gamma$ satisfy the relations
\[
\alpha' b^2 = \tfrac{1}{6}(D-26), \qquad b \ge 0,
\tag{34}
\]
\[
\sum_{n=1}^{D-1} p_n^2 = 1,
\qquad
\sum_{n=1}^{D-1} p_n = 1 + 2\gamma .
\tag{35}
\]
We have chosen $D>26$ and $0<t<\infty$. The solution for $D<26$ is obtained
by the substitution $b\rightarrow ib$, while the solution for $D=26$ is
found by taking the limit $b\rightarrow 0$ (and redefining $c_n$ and
$\phi_0$). Note that the form of the Minkowski signature solution with
$D>26$ ($D<26$) is identical to that of the Euclidean signature one
($t=-i\tau$) with $D<26$ ($D>26$). 

The space-time can be interpreted as a
product of the time line and the torus with time-dependent radii $a_n$.
Comparing (32), (33) with the model (6), (7) discussed above we can identify
one of the space coordinates $x_n$, e.g.\ $x_1$, with $y$ and the rest,
together with $t$, with $x^i$. Then it is easy to see that the solution (32)--(35)  
is indeed invariant under the duality transformation (19)
\[
\tilde a_1 = a_1^{-1}, \qquad
\tilde\phi = \phi - \ln a_1,
\tag{36}
\]
\[
\tilde p_1 = -p_1, \qquad
\tilde\gamma = \gamma - p_1, \qquad
a_1 = e^\lambda, \qquad
\lambda = p_1 \ln \tanh bt + \lambda_0 .
\]
The solution (32)--(35) is actually invariant under the $(D-1)$-dimensional
generalization of the duality (19)
\[
\tilde a_n(t) = a_n^{-1}(t), \qquad
\tilde\phi(t) = \phi(t) - \sum_{n=1}^{D-1}\ln a_n(t)
= \phi - \frac{1}{2}\ln \hat{G}, \qquad
\tilde\gamma = \gamma - \sum_{n=1}^{D-1} p_n .
\tag{37}
\]
More precisely, given a solution (32), (33) which satisfies the conditions
(35) there exists another one (37) with the inverse radii and shifted
dilaton. The simplest example is provided by the $D=2$ case of (32)--(35),
i.e., by the following (leading order) Euclidean solution of $D=2$ critical
string theory\foot{This solution, contained as a particular case in the solution
(32)--(35) of Ref.~18, was recently discussed (from various points of
view) in Ref.~19.} 
\[
ds^2 = d\tau ^2 + c_1^2 \tanh^2 b\tau \, dy^2, \qquad
\phi = \phi_0 - \ln \cosh b\tau,
\tag{38}
\]
\[
p_1 = 1, \qquad \gamma = 0, \qquad \alpha' b^2 = 4, \qquad \tau = it .
\]
The dual solution (36) is given by
\[
d\tilde s^2 = dr^2 + c_1^{-2} \coth^2 b\tau \, dy^2, \qquad
\tilde\phi = \tilde\phi_0 - \ln \sinh b\tau,
\tag{39}
\]
\[
p_1 = -1, \qquad \gamma = -1 .
\]
We see that the duality transformation may relate a regular metric (38)
to a singular one (39).\foot{Since the solutions (32)--(35), (38), (39) satisfy the string equations
of motion to the leading order in the $\alpha'$-expansion and hence are
valid only for large $t$ (or $\tau$) it is not clear whether the $\tau=0$
singularity is actually present in the exact solution which generalizes
(39).   }

The solution (32)--(35) with $D>26$ corresponds to the radii $a_n(t)$ with
$p_n>0$ ($p_n<0$) increasing (decreasing) to the constant values $c_n$
as $t\rightarrow\infty$, while $\phi(t)\rightarrow -bt$ so that the
effective string coupling $e^{\phi}$ decreases. The duality transformation
(36) reverses the behavior of $a_n(t)$ but does not change the asymptotic
form of the dilaton. A different picture is found for $D=26$. The $b\to0$
limit of (33) is  [18]
\[
a_n(t)=c_n' t^{p_n}, \qquad \phi(t)=\phi_0' + \gamma \ln t .
\tag{40}
\]
The conditions (35) have, for example, the following solution
\[
p_1=\frac12, \qquad
p_2=\frac12, \qquad
p_3=\frac{1}{\sqrt{2}}, \qquad
\gamma=\frac{1}{2\sqrt{2}}, \qquad
p_n=0, \qquad n=4,\ldots,25 .
\tag{41}
\]
describing three expanding dimensions and increasing string coupling.
The ``dual'' solutions (36), (37) correspond to changing the signs of any
number of $p_n$ and therefore have negative $\tilde{\gamma}$. Thus the
``non-static'' duality may in principle relate solutions with expansion
or contraction and increasing or decreasing effective string coupling
(see also Ref.~20).

As was noted in Ref.~18, the class of the solutions (32)--(35) has a
generalization to the second (and higher) order in $\alpha'$ expansion
which should therefore be invariant under the ``deformed'' duality
transformation (31). The transformation of the dilaton suggests a
connection with weak coupling--strong coupling duality. Another
interesting question is about the role of the ``non-static'' duality
and the dilaton in the qualitative picture of the dynamics of the
winding modes in string cosmology  [21].

\bigskip
\noindent
\textbf{Acknowledgments}

I am very grateful to M.~Rocek for drawing my attention to Ref.~10 and
emphasizing the question of higher-loop generalization of $\sigma$-model
duality transformation. I would also like to acknowledge a useful
conversation with G.~Veneziano.

After this paper was submitted for publication I learned that the
generalized (one-loop) duality transformation (4) was also discussed
in Ref.~22 where, in particular, the invariance of the leading term in
the effective action under (4) was pointed out.

The invariance of the solution (32),(33) under the duality
transformation (19) was first noted in Ref.~23. I would like to thank
F.~Quevedo for pointing this out to me and for a useful discussion of
some aspects of string duality.

\bigskip

\newpage
%\small 

\noindent
\textbf{References}

\begin{enumerate}
\item K.~Kikkawa and M.~Yamasaki, ``Casimir Effects in Superstring Theories'',  \emph{Phys.\ Lett.} B149 (1984) 357;

      N.~Sakai and I.~Senda, ``Vacuum Energies of String Compactified on Torus'',  \emph{Prog.\ Theor.\ Phys.} 75 (1986) 692.

\item V.~Nair, A.~Shapere, A.~Strominger, and F.~Wilczek, ``Compactification of the Twisted Heterotic String'', 
      \emph{Nucl.\ Phys.} B287 (1987) 402.

\item P.~Ginsparg and C.~Vafa, ``Toroidal Compactification of Nonsupersymmetric Heterotic Strings'',  \emph{Nucl.\ Phys.} B289 (1987) 414.

\item A.~Giveon, N.~Malkin, and E.~Rabinovici,  ``The Riemann Surface in the Target Space and Vice Versa'', 
      \emph{Phys.\ Lett.} B220 (1989) 551.

\item E.~Alvarez and M.~Osorio,  ``Duality Is an Exact Symmetry of String Perturbation Theory'',  \emph{Phys.\ Rev.} D40 (1989) 1150.

\item D.~Gross and I.~Klebanov, ``One-dimensional string theory on a circle'', 
\emph{Nucl.\ Phys.} B344 (1990) 475.

\item G.~Horowitz and A.~R.~Steif,  ``Is space-time duality violated in time dependent string solutions?'',  \emph{Phys.\ Lett.} B250 (1990) 49.

\item E.~Smith and J.~Polchinski,  ``Duality survives time dependence,''
Phys. Lett. B \emph{263} (1991)  59.
    %  Univ.\ Texas preprint UTTG-07

\item R.~Guven, ``Plane Waves in Effective Field Theories of Superstrings'', \emph{Phys.\ Lett.} B191 (1987) 275;

      D.~Amati and C.~Klimcik, ``Nonperturbative Computation of the Weyl Anomaly for a Class of Nontrivial Backgrounds'',  \emph{Phys.\ Lett.} B219 (1989) 443;
      
      G.~Horowitz and A.~R.~Steif,  ``Space-Time Singularities in String Theory'',  
       \emph{Phys.\ Rev.\ Lett.} 64 (1990) 260;
  ``Strings in Strong Gravitational Fields'', 
      \emph{Phys.\ Rev.} D42 (1990) 1950;
      
      R.~E.~Rudd, ``Compactification propagation'', 
       \emph{Nucl.\ Phys.} B352 (1991) 489.

\item T.~H.~Buscher, ``A Symmetry of the String Background Field Equations'',  \emph{Phys.\ Lett.} B194 (1987) 59;
     ``Path Integral Derivation of Quantum Duality in Nonlinear Sigma Models'',  
      \emph{Phys.\ Lett.} B201 (1988) 466.

\item A.~A.~Tseytlin, ``Partition function of string 
sigma  model on a compact two-space'',  \emph{Phys.\ Lett.} B223 (1989) 165;
   ``Sigma model approach to string theory'',    \emph{Int.\ J.\ Mod.\ Phys.} A4 (1989) 1257, [arXiv:2602.10977];
      
      O.~D.~Andreev, R.~R.~Metsaev, and A.~A.~Tseytlin,  ``Covariant calculation of the statistical sum of the two-dimensional sigma model on compact two surfaces.'', 
      \emph{Sov.\ J.\ Nucl.\ Phys.} 51 (1990) 359,  [arXiv:2301.02867].

\item S.~Randjbar-Daemi, A.~Salam, and J.~Strathdee, ``Sigma models and string theories'', 
      \emph{Int.\ J.\ Mod.\ Phys.} A2 (1987) 667;
      
      S.~de~Alwis, ``The Freedom of the {Wess-Zumino}-Witten Model'', \emph{Phys.\ Lett.} B164 (1985) 67.

\item M.~J.~Duff,  ``Observations on Conformal Anomalies'',  \emph{Nucl.\ Phys.} B125 (1977) 334;

      A.~M.~Polyakov, ``Quantum Geometry of Bosonic Strings'', \emph{Phys.\ Lett.} B103 (1981) 207.

\item E.~S.~Fradkin and A.~A.~Tseytlin, ``Effective Field Theory from Quantized Strings'', 
      \emph{Phys.\ Lett.} B158 (1985) 316;
      ``Quantum String Theory Effective Action'', 
      \emph{Nucl.\ Phys.} B261 (1985) 1.

\item D.~Friedan, ``Nonlinear Models in Two  Epsilon Dimensions'', \emph{Phys.\ Rev.\ Lett.} 45 (1980) 1057;
    ``Nonlinear Models in Two + Epsilon Dimensions'',   \emph{Ann.\ Phys.} 163 (1985) 318;

      C.~Callan, D.~Friedan, E.~Martinec, and M.~Perry,  ``Strings in Background Fields'', 
      \emph{Nucl.\ Phys.} B262 (1985) 593;
      
      C.~Callan, I.~Klebanov, and M.~Perry,  ``String Theory Effective Actions'', 
      \emph{Nucl.\ Phys.} B278 (1986) 78.

\item A.~A.~Tseytlin,  ``Vector Field Effective Action in the Open Superstring Theory'', 
 \emph{Nucl.\ Phys.} B276 (1986) 391;
     ``Conformal Anomaly in Two-Dimensional Sigma Model on Curved Background and Strings'',  \emph{Phys.\ Lett.} B178 (1986) 34.

\item A.~A.~Tseytlin, ``Conditions of Weyl Invariance of Two-dimensional 
Sigma  Model From Equations of Stationarity of 'Central Charge' Action'', \emph{Phys.\ Lett.} B194 (1987) 63;
   ``Mobius Infinity Subtraction and Effective Action in 
Sigma  Model Approach to Closed String Theory'',    \emph{Phys.\ Lett.} B208 (1988) 221;

      H.~Osborn,  ``String Theory Effective Actions From Bosonic 
Sigma  Models'',  \emph{Nucl.\ Phys.} B308 (1988) 629.

\item M.~Mueller, ``Rolling Radii and a Time Dependent Dilaton'', \emph{Nucl.\ Phys.} B337 (1990) 37.

\item S.~Elitzur, A.~Forge, and E.~Rabinovici, ``Some global aspects of string compactifications,''
 \emph{Nucl. Phys.} B{359} (1991)  58;

      M.~Rocek, K.~Schoutens, and A.~Sevrin, ``Off-shell WZW models in extended superspace,''
\emph{Phys. Lett.} B{265} (1991) 303; 
     % IAS preprint IASSNS-HEP-91/14;
      
      E.~Witten, ``On string theory and black holes,''
\emph{Phys. Rev.} D44 (1991) 314; %   IAS preprint IASSNS-HEP-91/12;
      
      G.~Mandal, A.~M.~Sengupta, and S.~R.~Wadia, ``Classical solutions of two-dimensional string theory,''
      \emph{Mod.\ Phys.\ Lett.} A6 (1991) 1685.

\item G.~Veneziano, ``Scale factor duality for classical and quantum strings,''
\emph{Phys. Lett.} B{265} (1991)  287.

\item R.~Brandenberger and C.~Vafa,  ``Superstrings in the Early Universe'', 
      \emph{Nucl.\ Phys.} B316 (1988) 391.

\item T.~Banks, M.~Dine, H.~Dijkstra, and W.~Fischler,  ``Magnetic Monopole Solutions of String Theory'', 
      \emph{Phys.\ Lett.} B212 (1988) 45.

\item L.~Ibanez, D.~Lust, F.~Quevedo, and S.~Theisen,
      unpublished (1990).

\end{enumerate}

\end{document}